\documentclass{caosp}
\usepackage{graphicx}
\usepackage{url}

\articleNo{225}
\pubyear{2013}
\volume{43}
\volnumber{1}
\firstpage{5}
\received{January 28, 2013}
\accepted{February 20, 2013}

\newcommand{\ha}{H$\alpha$}
\newcommand{\fw}{{\it FW}}
\newcommand{\ic}{$I_{\rm C}$}
\newcommand{\vc}{$v_{\rm C}$}
\newcommand{\vbi}{$v_{\rm BI}$}
\newcommand{\pap}{Paper\,I}
\newcommand{\kms}{\,km\,s$^{-1}$}

\newcommand{\ifw}{$I_{\rm FW}$}

\begin{document}

\hauthor{J.\,Koza, P.\,S\"{u}tterlin, P.\,G\"{o}m\"{o}ry, J.\,Ryb\'{a}k and A.\,Ku\v{c}era}

\title{Search for Alfv\'{e}n waves in a bright network element observed in H$\alpha$}

\author{
  J.\,Koza\inst{1}
  \and
  P.\,S\"{u}tterlin\inst{2}
  \and
  P.\,G\"{o}m\"{o}ry\inst{1}
  \and
  J.\,Ryb\'{a}k\inst{1}
  \and
  A.\,Ku\v{c}era\inst{1}
}

\institute{
  \lomnica, \email{koza@astro.sk}
  \and
  Institute for Solar Physics, The Royal Swedish Academy of Sciences, Alba Nova University Center, 106 91 Stockholm, Sweden
}

\date{January 28, 2013}

\maketitle

\begin{abstract}
  Alfv\'{e}n waves are considered as potential transporters of energy
  heating the solar corona. We seek spectroscopic signatures of the
  Alfv\'{e}n waves in the chromosphere occupied by a bright network
  element, investigating temporal variations of the spectral width,
  intensity, Dopplershift, and the asymmetry of the core of the
  \ha\ spectral line observed by the tunable Lyot filter installed on
  the Dutch Open Telescope. The spectral characteristics are derived
  through the fitting of five intensity samples, separated from each
  other by 0.35\,\AA, by a $4^{\rm th}$-order polynomial. The bright
  network element displays the most pronounced variations of the
  Dopplershift varying from 0 to 4\kms\ about the average of
  1.5\kms. This fact implies a persistent redshift of the \ha\ core
  with a redward asymmetry of about 0.5\kms, suggesting an inverse-C
  bisector. The variations of the core intensity up to $\pm 10$\,\%
  and the core width up to $\pm 5$\,\% about the respective averages
  are much less pronounced, but still detectable. The core intensity
  variations lag behind the Dopplershift variations about
  2.1\,min. The \ha\ core width tends to correlate with the
  Dopplershift and anticorrelate with the asymmetry, suggesting that
  more redshifted \ha\ profiles are wider and the broadening of the
  \ha\ core is accompanied with a change of the core asymmetry from
  redward to blueward. We also found a striking anticorrelation
  between the core asymmetry and the Dopplershift, suggesting a change
  of the core asymmetry from redward to blueward with an increasing
  redshift of the \ha\ core. The data and the applied analysis do not
  show meaningful tracks of Alfv\'{e}n waves in the selected network
  element.
\end{abstract}

\section{Introduction}

Alfv\'{e}n waves have been invoked as one of possible mechanisms
heating the solar corona (Alfv\'{e}n, 1947; Osterbrock, 1961). These
magnetic transporters of convective energy are capable of penetrating
through the solar atmosphere without being reflected and
refracted. Since Alfv\'{e}n waves are incompressible, their
propagation through the atmosphere is not associated with density
changes seen as periodic variations of intensities and line-of-sight
velocities (Erd\'{e}lyi \& Fedun, 2007; Mathioudakis {\it et al.},
2012). Observation of a slanted magnetic fluxtube could reveal
Alfv\'{e}n waves as periodic variations of nonthermal broadening of a
spectral line. This approach was employed in the study by Jess {\it et
  al.} (2009) who found periodic oscillations of nonthermal broadening
of the \ha\ spectral line seen over a cluster of bright points
associated with a distinct upflow without significant periodic
variations of intensities and line-of-sight velocities. They interpret
this as a manifestation of Alfv\'{e}n waves in the solar
atmosphere. However, there is some ambiguity whether they deal with
spectral or integrated intensities.

Motivated by the results of Jess {\it et al.} (2009), we search for
signatures of Alfv\'{e}n waves in a bright network element occupied by
a cluster of G-band bright points.  We investigate spectral
characteristics of the \ha\ spectral line, focusing on temporal
variations of the line core width, intensity, Dopplershift, and the
asymmetry derived by a $4^{\rm th}$-order-polynomial fitting of a
coarse five-point sampling of the \ha\ profile. Since we employ the
correcting method and results of Koza {\it et al.} (2013), and
frequently refer to it as \pap\ hereafter, we invite the reader to
have it handy.

\begin{figure}
  \centering
  \includegraphics[width=0.84\textwidth,clip=]{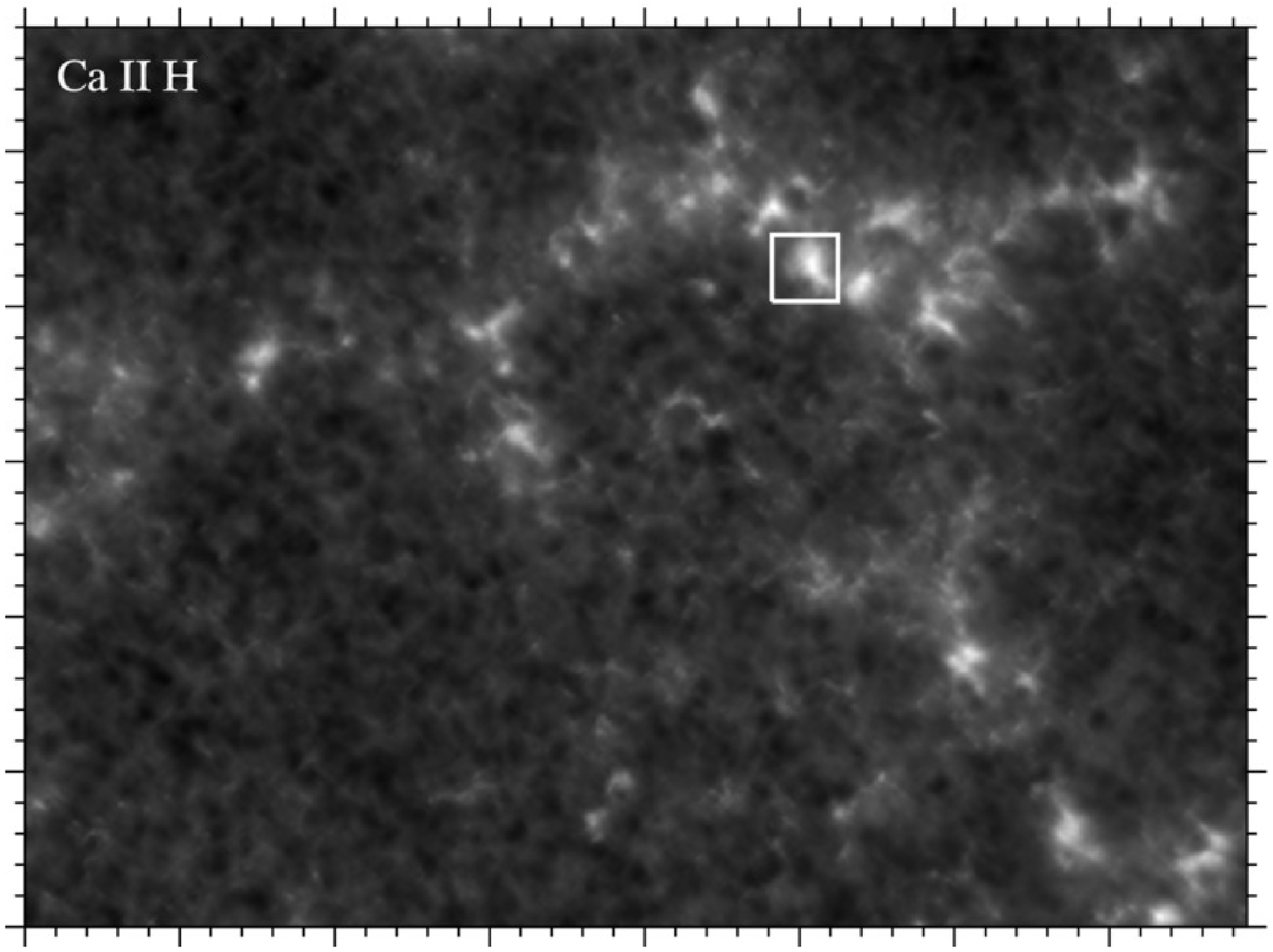} \\
  \includegraphics[width=0.84\textwidth,clip=]{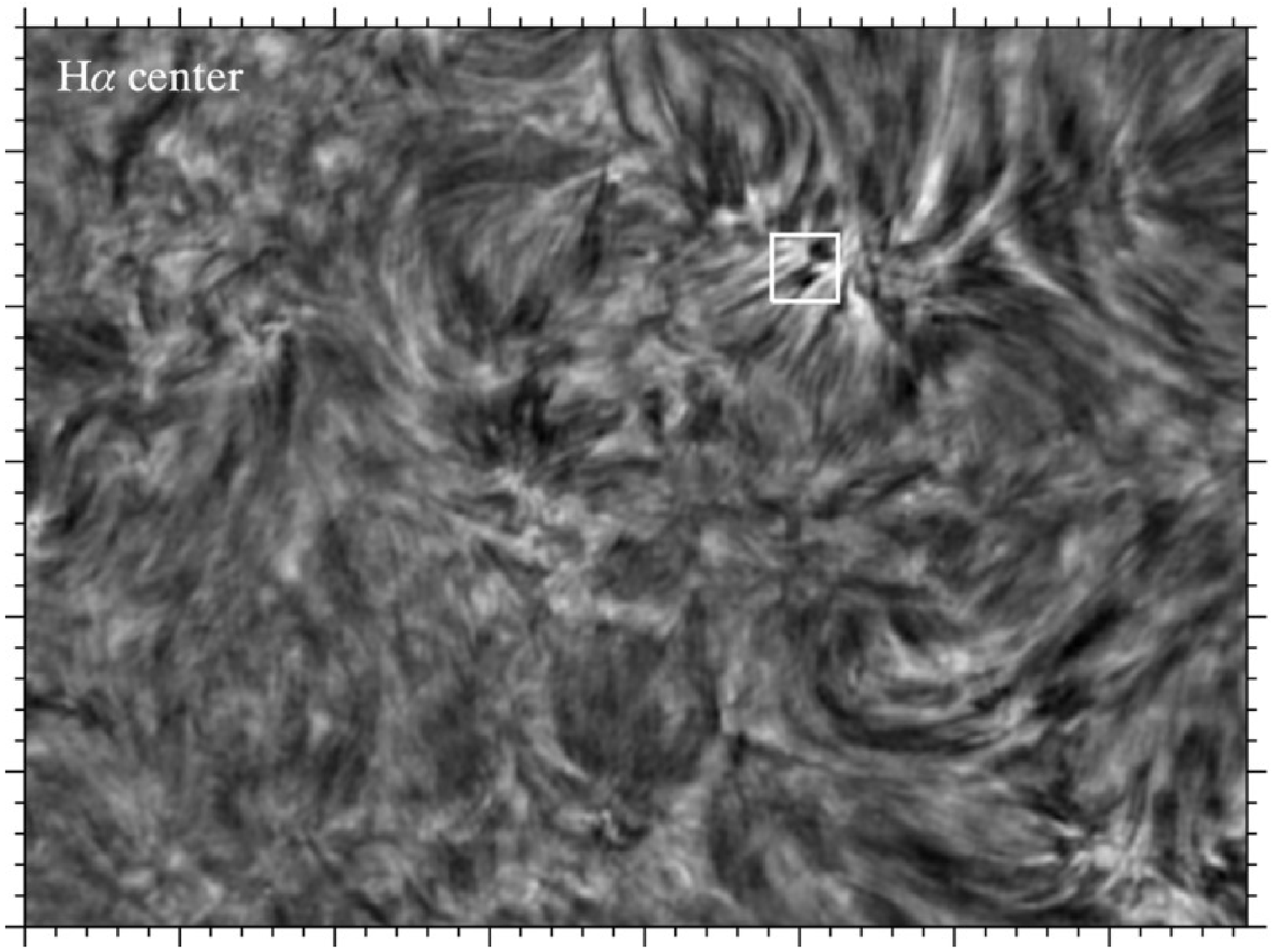}
  \caption{Images of the quiet-Sun network recorded by DOT on 19 October
    2005.  {\it Top:} the temporal mean of the Ca\,II\,H image sequence
    over its 71-min duration.  {\it Bottom:} the \ha\ image taken at
    10:36:21\,UT in the line center at the moment of the best seeing
    occurred in the $41^{\rm st}$\,min after the start of the
    observation at 09:55:20\,UT. The square defines the network element
    selected for this study and displayed in detail in
    Figs.\,\ref{fig4}--\ref{fig7}. Field of view: 79 $\times$
    58\,arcsec$^2$. Tickmark spacing: 2\,arcsec.}
  \label{fig1}
\end{figure}

\begin{figure}
  \centering
  \includegraphics[width=0.84\textwidth,clip=]{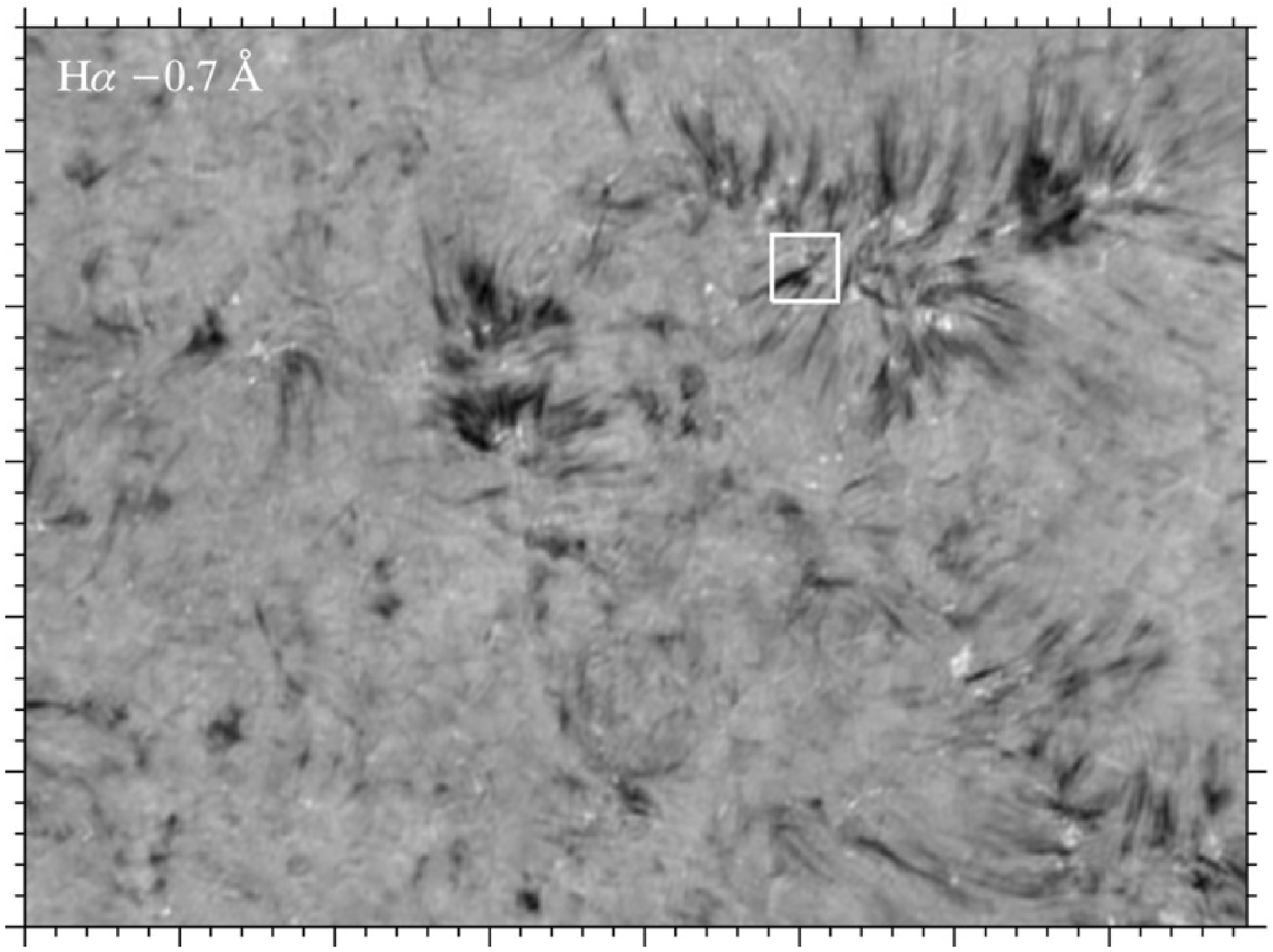} \\
  \includegraphics[width=0.84\textwidth,clip=]{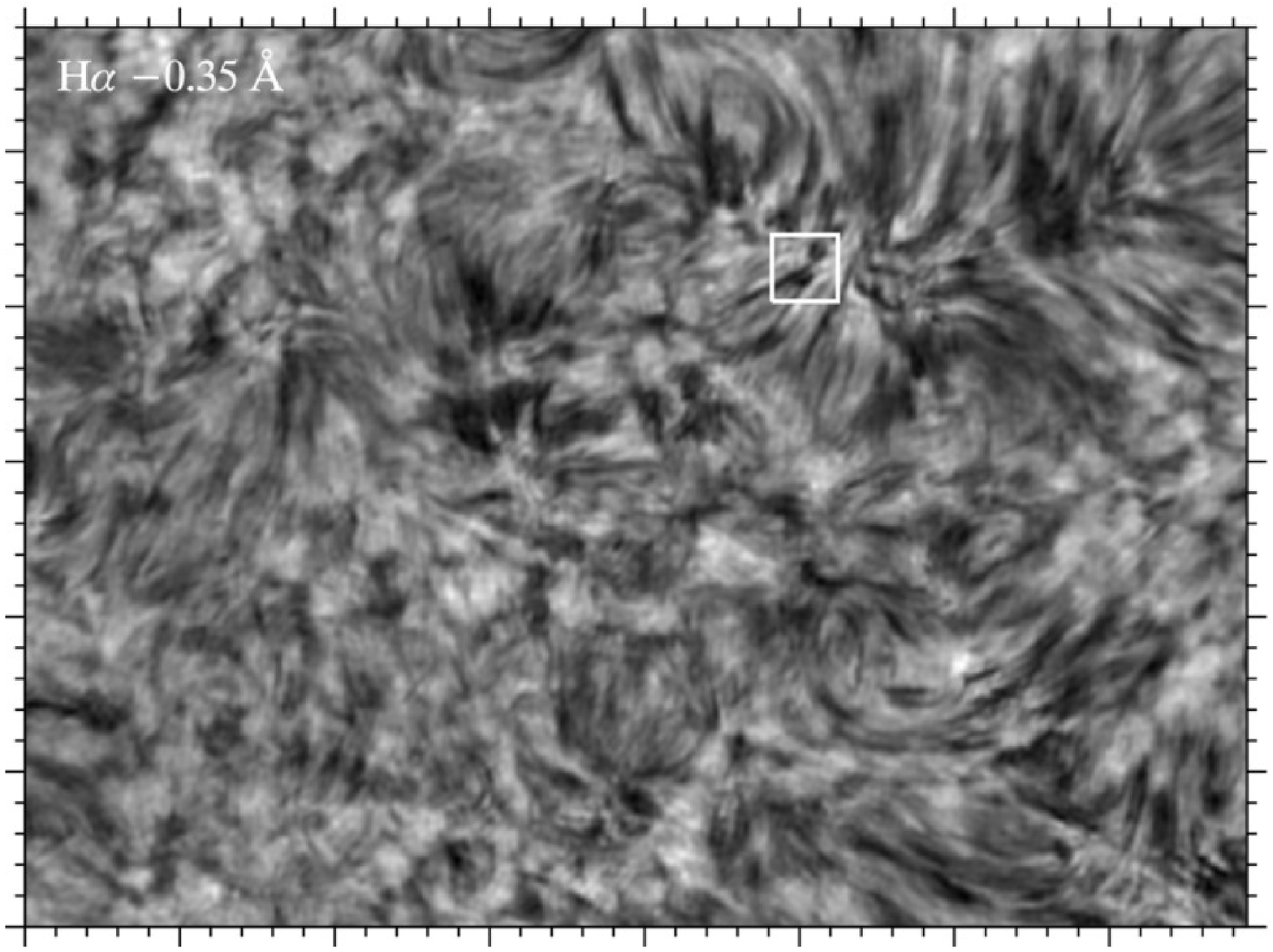}
  \caption{DOT \ha\ images at $\Delta\lambda = -0.7$\,\AA\ ({\it top})
    and $-0.35$\,\AA\ ({\it bottom}) from the line center
    corresponding to the \ha\ line-center image in Fig.\,\ref{fig1}.}
  \label{fig2}
\end{figure}

\begin{figure}
  \centering
  \includegraphics[width=0.84\textwidth,clip=]{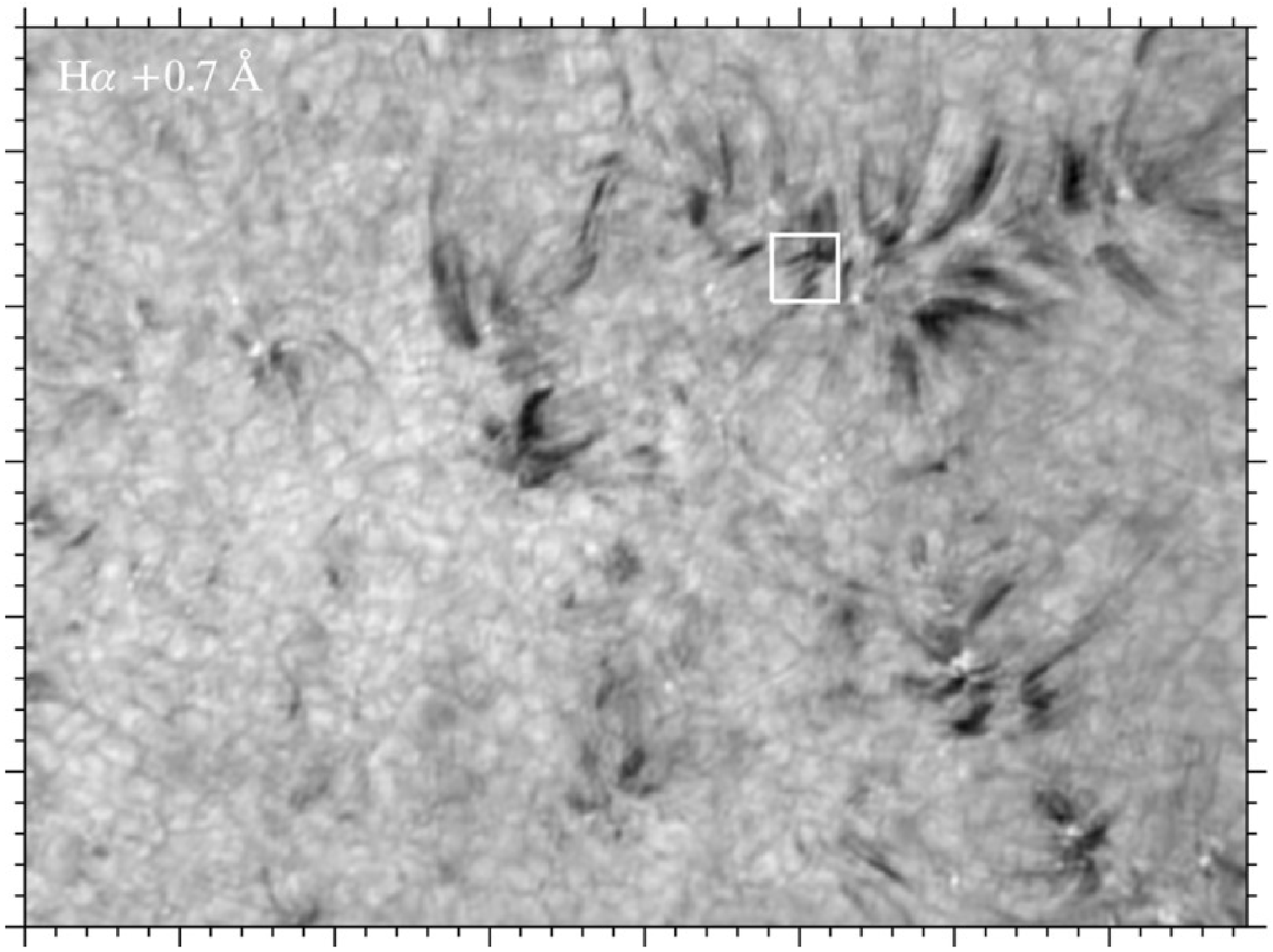} \\
  \includegraphics[width=0.84\textwidth,clip=]{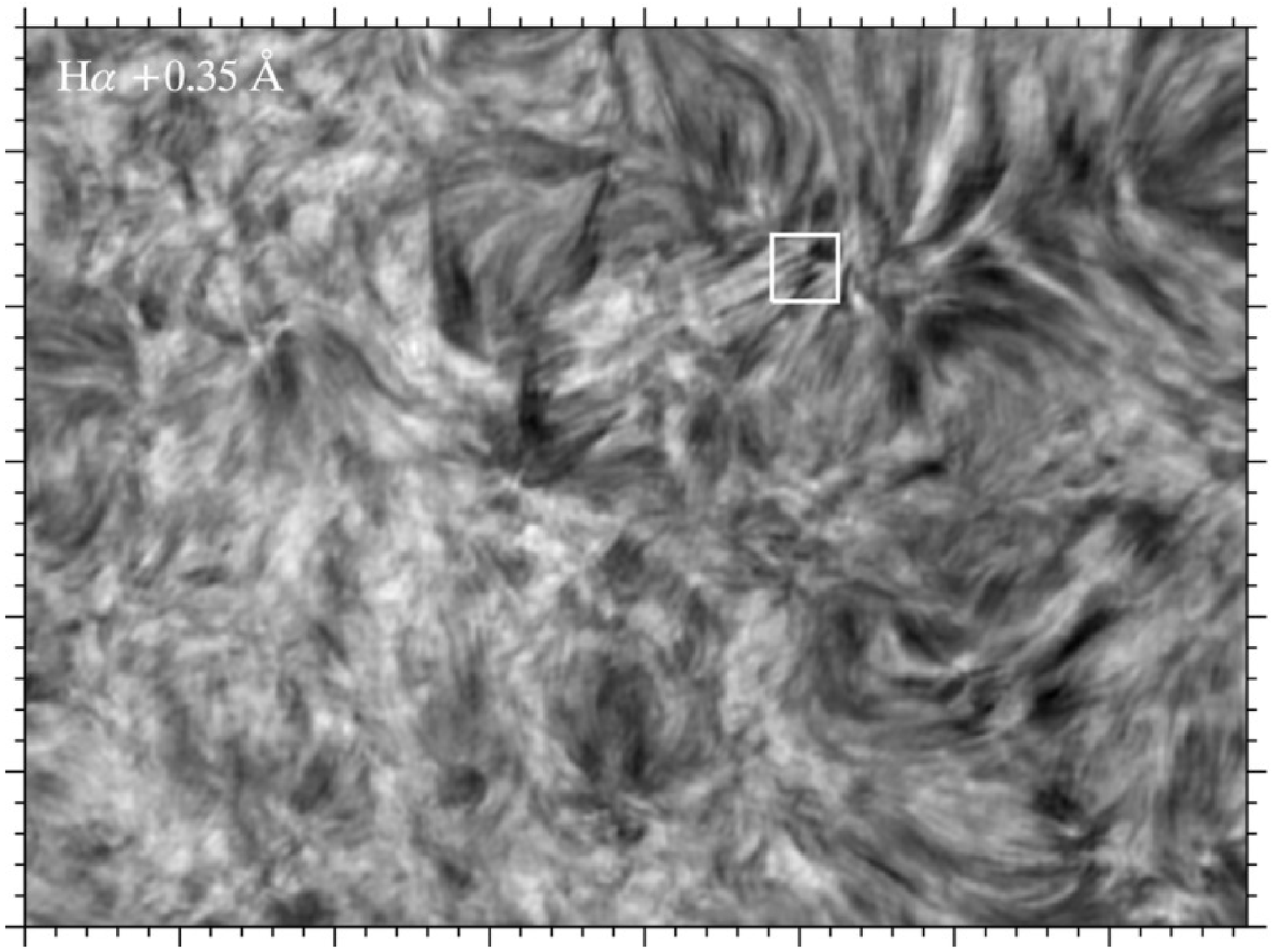}
  \caption{DOT \ha\ images at $\Delta\lambda = +0.7$\,\AA\ ({\it top})
    and +0.35\,\AA\ ({\it bottom}) from the line center corresponding
    to the \ha\ images in Figs.\,\ref{fig1} and \ref{fig2}.}
  \label{fig3}
\end{figure}

\begin{figure}
  \centering
  \includegraphics[width=0.83\textwidth,clip=]{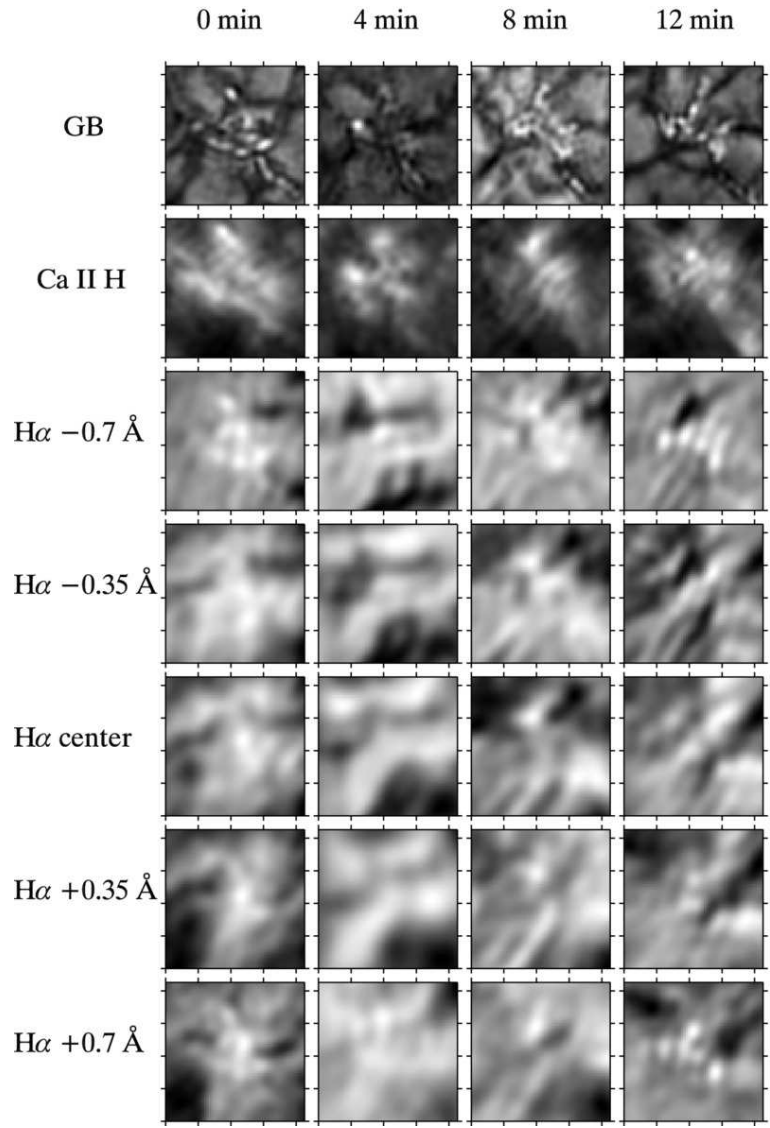}
  \caption{Temporal evolution of the selected network element seen
    simultaneously in the DOT G-band, Ca\,II\,H, and \ha\ images taken
    at five wavelength settings of the \ha\ Lyot filter. The time 0\,min
    corresponds to the moment of the start of the observation at
    09:55:20\,UT. Field of view: 4.3 $\times$ 4.3\,arcsec$^2$. Tickmark
    spacing: 1\,arcsec.}
  \label{fig4}
\end{figure}

\begin{figure}
  \centering
  \includegraphics[width=0.83\textwidth,clip=]{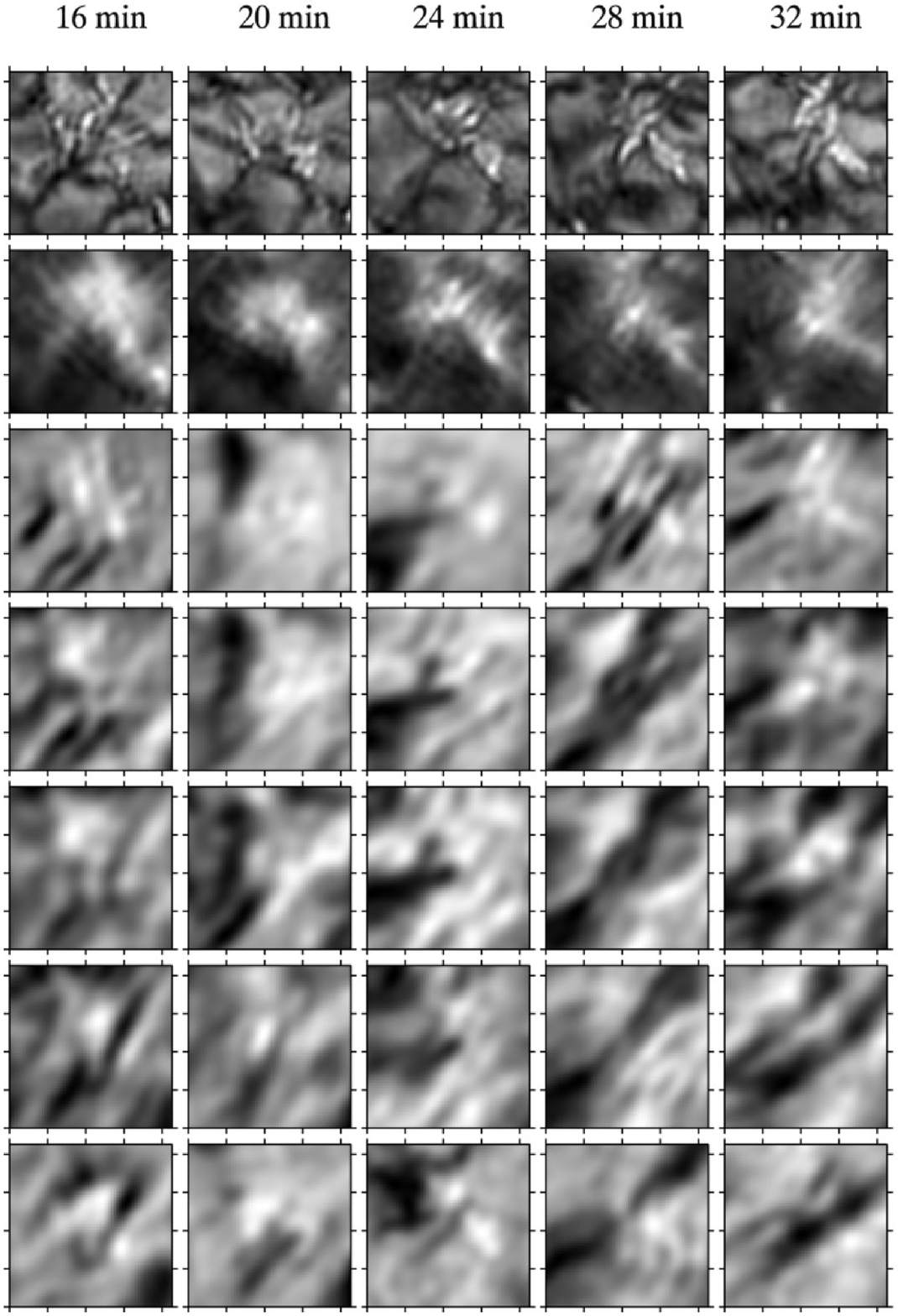}
  \caption{The same as in Fig.\,\ref{fig4}.}
  \label{fig5}
\end{figure}

\begin{figure}
  \centering
  \includegraphics[width=0.83\textwidth,clip=]{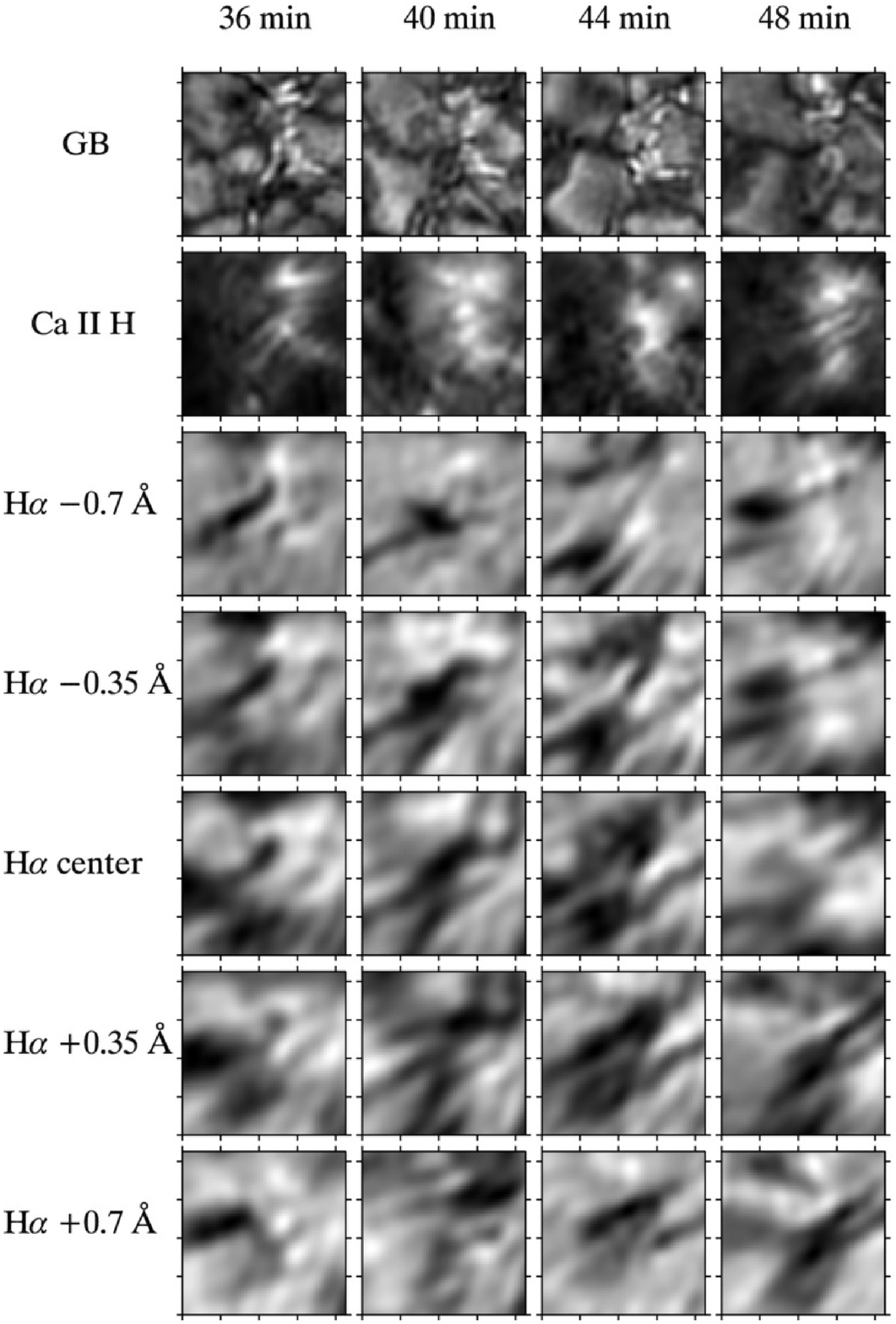}
  \caption{The same as in Fig.\,\ref{fig4}.}
  \label{fig6}
\end{figure}

\begin{figure}
  \centering
  \includegraphics[width=0.83\textwidth,clip=]{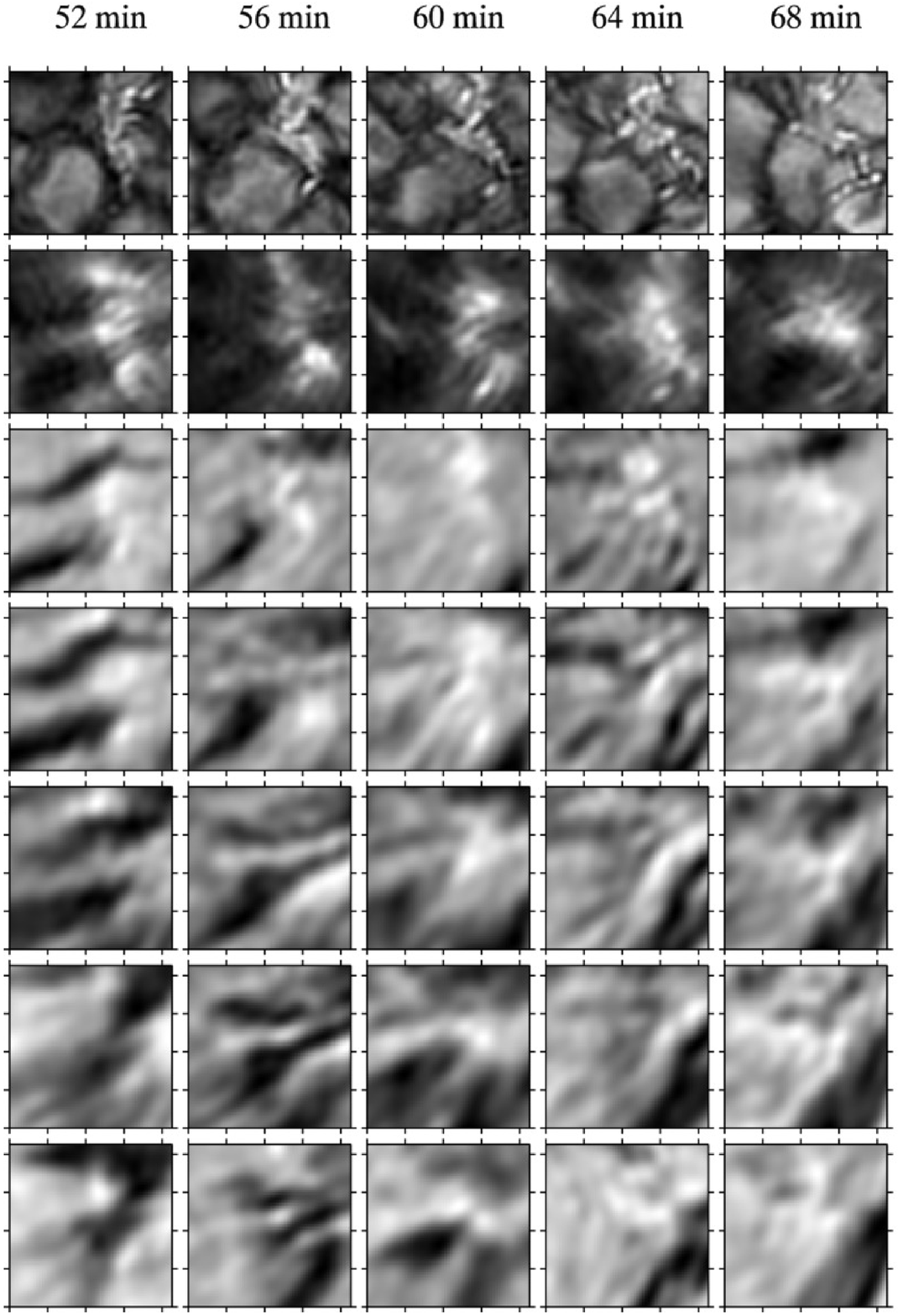}
  \caption{The same as in Fig.\,\ref{fig4}.}
  \label{fig7}
\end{figure}

\section{Observations}
\label{obs}

We use sequences of \ha\ images of a very quiet area near the disk
center ($\mu = 0.91$) recorded by the Dutch Open Telescope (DOT) on 19
October 2005 from 09:55:20\,UT to 11:05:47\,UT by the tunable Lyot
filter with a {\it FWHM} passband of 0.25\,\AA\ (Gaizauskas, 1976;
Bettonvil {\it et al.}, 2006). The sequence consists of scans taken
in five wavelengths across the \ha\ line profile: line center, $\pm
0.35$\,\AA\ and $\pm 0.7$\,\AA, followed by a single image in the line
center. The zero wavelength of the filtergrams is centered on the
minimum of an average profile taken before the observations over the
field of view. A time step between the five-wavelength scans of the
\ha\ line profile and line center images is 30\,s. Since the images at
two subsequent wavelengths are separated by the 4-to-5-s intervals,
one full 5-point scan of the \ha\ profile took about 20-25\,s.  Within
this interval no major changes of chromospheric scenery are
assumed. The scans were taken in 20-frame bursts per particular
wavelength setting and then speckle-reconstructed applying a Keller -
von der L\"uhe two-channel reconstruction (Keller \& von der L\"uhe,
1992). The line center images were obtained in 100-frame bursts
restored by a full single-channel speckle reconstruction. In this
study, we employ 71 \ha\ scans at a regular 60-s cadence taken between
09:55:20\,UT and 11:05:22\,UT. The seeing was only fair, with the
Fried parameter $r_0$ slightly increasing from about 7.5 at the
beginning to 9 on average at the end of the observation. As a context
data, we also use the sequences of 142 G-band and Ca\,II\,H images
taken synchronously with the \ha\ images at a regular 30-s
cadence. The temporal mean of the Ca\,II\,H images and examples of the
\ha\ images taken during the period of the best seeing are shown in
Figs.\,\ref{fig1}--\ref{fig3}. The size of the Ca\,II\,H and
\ha\ images is $1112 \times 818$\,px$^2$, with the angular size of
pixel of 0.071\,arcsec. Details on the DOT, its tomographic
multiwavelength imaging, speckle reconstruction, and standard
reduction procedures are given in Hammerschlag \& Bettonvil (1998) and
Rutten {\it et al.}  (2004).

The target area encompasses a full supergranular cell surrounded by a
prominent network as seen in the temporal mean of the Ca\,II\,H
sequence in Fig.\,\ref{fig1}. Visual comparison of both panels in
Fig.\,\ref{fig1} indicates that the brightest network elements
coincide with roots of long chromospheric fibrils. These dominate
\ha\ scene and emanate out from the network elements outlining the
structure of magnetic field at chromospheric levels. DOT \ha\ scans
render a tomographic view of the solar atmosphere from the photosphere
(Figs.\,\ref{fig2} and \ref{fig3}: $\pm 0.7$\,\AA) up to the
chromosphere sampled by the images in the line center and $\pm
0.35$\,\AA\ off (Kontogiannis {\it et al.}, 2010). In the \ha\ $+
0.7$\,\AA\ image (Fig.\,\ref{fig3}, top), we can clearly see
granulation which is missing in the \ha\ $- 0.7$\,\AA\ image
(Fig.\,\ref{fig2}, top) due to Doppler cancellation. Both images
display dark elongated streaks which are highly-dopplershifted parts
of fibrils seen better in the line center and $\pm
0.35$\,\AA\ images. Doppler cancellation enhances the visibility of
magnetic elements seen as bright points in the \ha\ $-
0.7$\,\AA\ image (Fig.\,\ref{fig2}, top), studied in detail in
Leenaarts {\it et al.}  (2006). The square defines the network element
selected for this study. The size of the square is $61 \times
61$\,px$^2$, corresponding to $4.3 \times 4.3$\,arcsec$^2$.

Figs.\,\ref{fig4}--\ref{fig7} show enlargements of the square
providing a multispectral tomographic view on evolution of the
selected network element unfolded in time with a 4-min timestep from
the lower photosphere (G band) up to the chromosphere (\ha\ line
center). The G-band cutouts show a cluster of bright points
constituting the network element. The selected area also involves two
large granules, with maximum diameters larger than 2\,arcsec,
occurring in the lower half from 16 to 28\,min (Fig.\,\ref{fig5}) and
from 44 to 68\,min (Figs.\,\ref{fig6} and \ref{fig7}). The Ca\,II\,H
cutouts show brightenings whose overall morphology is often very
similar to that of the cluster of the G-band bright points. But the
former are strongly diffuse because of resonance scattering within the
solar atmosphere and through expansion of idealized magnetostatic
fluxtubes with height (Leenaarts {\it et al.}, 2006). The \ha\ outer
wing cutouts ($\pm 0.7$\,\AA) mix photospheric and chromospheric
features because of a double-peak contribution function at \ha\ wings
(Schoolman, 1972; Leenaarts {\it et al.}, 2006, 2012). Striking,
almost one-to-one correspondence between the G-band bright points and
the blue wing bright points seen in $- 0.7$\,\AA\ (compare the G-band
and \ha\ $- 0.7$\,\AA\ cutouts, {\it e.g.}, at 0, 12, 32, 60, and
64\,min) was explained in Leenaarts {\it et al.}  (2006). The line
center and $\pm 0.35$\,\AA\ cutouts bear still signatures of a network
brightening, but modulated much with a fast-changing blanket of the
dark fibrils occurring higher up.

\section{Method, spectral characteristics, and their corrections}
\label{datanal}

The coarse five-point sampling enables us to construct $1112 \times
818 \times 71$ instantaneous proxy profiles of the \ha\ spectral line
at each pixel.  We fit the five-wavelength samples of the proxy
profiles by a $4^{\rm th}$-order polynomial to derive the four basic
profile measurements: the core intensity \ic, the core velocity \vc,
the core width \fw, and the bisector velocity \vbi. The values of
\ic\ and \vc\ represent the fit minimum and its Dopplershift,
respectively. The core width is the wavelength separation of the two
fit flanks at half of the intensity range between the fit minimum and
the average of the endpoint intensities at $\pm 0.7$\,\AA. The core
width is computed at the average intensity \ifw\ employing the
reference intensity $I_{\rm RF1}$ defined by equations~(3) and (4) in
\pap. Finally, the bisector velocity is the velocity equivalent of the
wavelength separation between the midpoint of the fit at the average
intensity \ifw\ and the fit minimum. A more detailed definition of the
spectral characteristics is given in \pap.  In case of a large
Dopplershift, when the average intensity is larger than the minimum of
endpoint intensities, {\it i.e.}, $I_{\rm FW} > {\rm min}(I_{\rm
  -0.7}, I_{\rm +0.7})$, the fits are extrapolated beyond the range
$\pm 0.7$\,\AA\ using the polynomial coefficients and the extrapolated
fits are used for determining the spectral characteristics. In very
rare cases of a polynomial undulation, with local maxima lower than
the average intensity \ifw, the algorithm switches to a parabolic fit
of the five-point proxy profile to determine \fw, while \ic\ and
\vc\ are always from the $4^{\rm th}$-order-polynomial fit and
\vbi\ is undefined. The extrapolation (Fig.\,\ref{fig8}) was invoked
just in 0.024\,\% and the parabolic fit in $5 \times 10^{-4}$\,\% out
of all pixels.  The intensities shown in Figs.\,\ref{fig8} and
\ref{fig9} were normalized with respect to the mean intensities
resulting from spatio-temporal averaging of the respective image
cubes. The five mean intensities were converted to the atlas intensity
scale convolving the \ha\ atlas profile with the transmission profile
of the DOT \ha\ filter (\pap, equation~(1)). Therefore, the integrated
intensities in Figs.\,\ref{fig8} and \ref{fig9} are in the wavelength
unit.

The five intensity samples have an instrumental wavelength scale
ranging from $-0.7$ to 0.7\,\AA. The wavelengths of particular core
fit minima $\lambda_{\rm C}$ are measured with respect to this
instrumental wavelength scale with milli{\aa}ngstr\"om resolution and
then converted to the core velocity as: $v_{\rm C} = c(\lambda_{\rm C}
- \lambda_{\rm R})/\lambda_0$\,, where $c$ is the speed of light,
$\lambda_0$ is the central wavelength of the \ha\ line 6562.8\,\AA,
and the reference wavelength $\lambda_{\rm R}$ is the spatio-temporal
average of all $1112 \times 818 \times 71$ wavelengths of fit minima
$\lambda_{\rm C}$ over all pixels within the field of view, making no
distinction between network and internetwork pixels. The reference
wavelength $\lambda_{\rm R}$ is about $-18$\,m\AA. An adopted sign
convention indicates a redshift of the line core fit for positive
Dopplershift and \vc.
The bisector velocity \vbi\ quantifies an asymmetry of the line core fit with
respect to its minimum. The sign convention adopted in \pap\ suggests
positive \vbi\ for a line core fit with a redward asymmetry.

\begin{figure}
  \centering
  \includegraphics[width=0.76\textwidth, clip=]{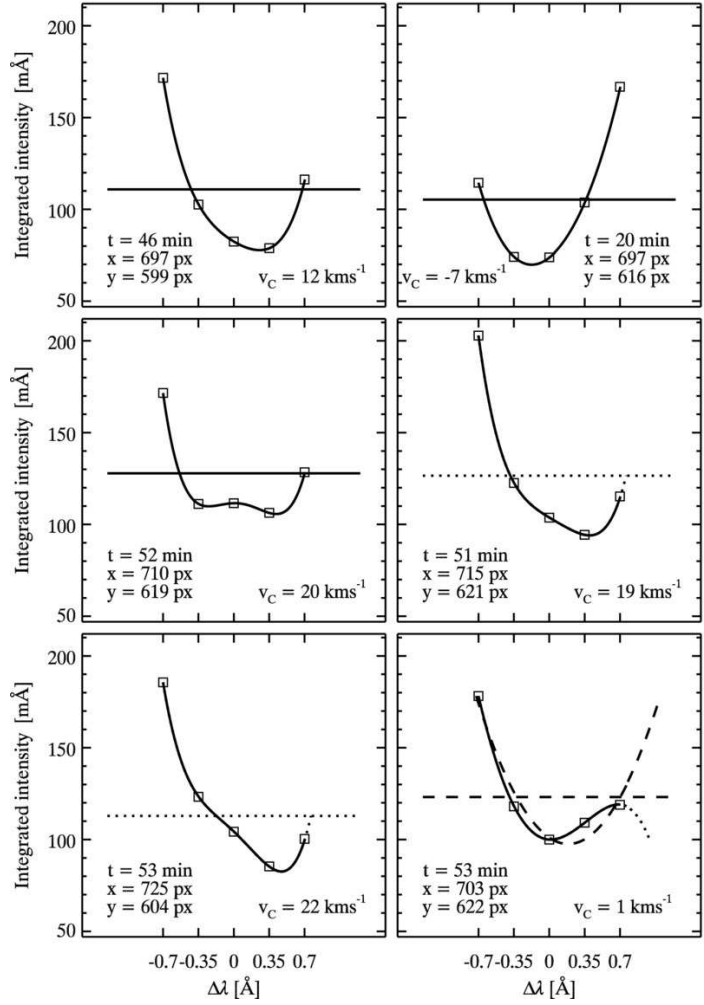}
  \caption{Examples of observed intensities (squares), their $4^{\rm
      th}$-order-polynomial (solid), extrapolated (dotted), and
    parabolic (dashed) fits taken from various locations $x, y$ over
    the selected network element at the indicated time $t$. The
    horizontal lines indicate the average intensity defining the fit
    width and the bisector velocity. Their solid, dotted, and dashed
    linestyles correspond to the $4^{\rm th}$-order-polynomial,
    extrapolated, and parabolic fits, respectively, used in
    determination of the fit width and the bisector velocity. The
    coordinates $x, y$ refer to the lower left corners of images. The
    selected samples do not represent majority, but rather rare cases
    with a high core velocity \vc\ and highly asymmetric profiles (the
    middle and bottom panels).}
  \label{fig8}
\end{figure}

\begin{figure}
  \centering
  \includegraphics[width=0.86\textwidth, clip=]{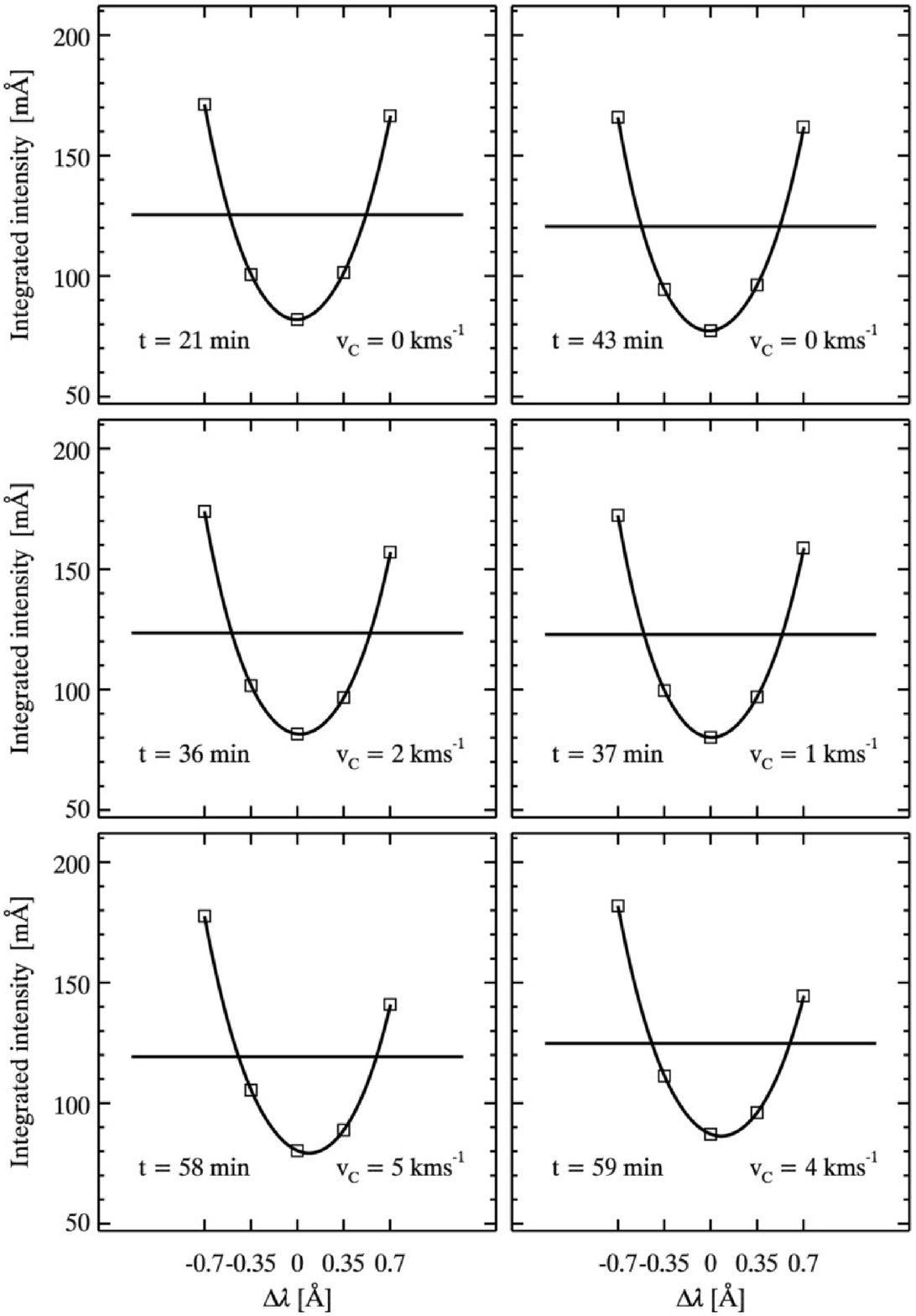}
  \caption{Examples of observed intensities (squares) averaged over the
    selected network element (Figs.\,\ref{fig4}--\ref{fig7}) and their
    $4^{\rm th}$-order-polynomial fits (solid) at the given time
    $t$. The horizontal lines indicate the average intensity defining
    the fit width and the bisector velocity. The samples illustrate the
    fits with small (top), medium (middle), and high (bottom) redshift
    with the core velocity \vc.}
  \label{fig9}
\end{figure}

To eliminate the modulation of the spectral characteristics induced by
the fast-changing blanket of dark fibrils, we construct 71 proxy
\ha\ profiles by spatial averaging of intensities within the selected
network element (Figs.\,\ref{fig4}--\ref{fig7}) at each
wavelength. Fig.\,\ref{fig9} shows examples of these spatially
averaged intensities, their polynomial fits, average intensities, and
core velocities. Spectral characteristics of these spatially averaged
intensities are determined solely from their $4^{\rm
  th}$-order-polynomial non-extrapolated fits. To asses an influence
of the averaging on the final results, we compare them with the
instantaneous characteristics obtained by the spatial averaging of
\fw, \ic, \vc, and \vbi\ within the selected area
(Figs.\,\ref{fig4}--\ref{fig7}).
Both, the instantaneous spectral characteristics inferred per pixel
and also from the spatially averaged intensities are corrected for the
deviations estimated in \pap\ 
comparing the measurements of \fw, \ic, \vc, and \vbi\ performed at
the dopplershifted \ha\ atlas profile convolved with the filter
transmission with their reference values. Then the corrected core
velocity \vc\ is the solution of the equation $f(v_{\rm C}) + v_{\rm
  C} - \nu = 0$\,, where $f(v_{\rm C})$ is the absolute deviation
$\Delta v$ and $\nu$ is the measured core velocity. The corrected
\fw\ results from the equation $FW = fw - \Delta FW(v_{\rm C})$\,,
where $fw$ is the measured fit width and $\Delta FW(v_{\rm C})$ is the
absolute deviation $\Delta FW$, but interpolated on the corrected core
velocity \vc.  The corrected \vbi\ is computed in the same way. Due to
different intensity scales, the corrected \ic\ results from the
equation $I_{\rm C} = i/(1 + \Delta I_{\rm r}(v_{\rm C})/100)$\,,
where $i$ is the measured core intensity and $\Delta I_{\rm r}(v_{\rm
  C})$ is the relative deviation $\Delta I_{\rm r}$, but interpolated
on the corrected core velocity \vc.

\section{Noise, data uncertainties, and references}
\label{errors}

Reliability of results inferred by fitting depends critically on the
noise presented in the data and various intrinsic
uncertainties. Therefore, in this section we give a brief account of
uncertainties of the fitted intensities and the reference wavelength
defining the core velocity.

We estimated the upper limit of the noise level $1\sigma_{\rm NOISE}$
presented in the \ha\ filtergrams from $1\sigma$ of the temporal
variations of the intensity at each pixel assuming that $1\sigma_{\rm
  NOISE} \leq 1\sigma/10$. In other words, we assumed that the noise
level is about one order smaller than $1\sigma$ of the temporal
variations of the intensity due to physical processes on the Sun
within the 71-min observing period. This rough estimate points out that
these $1\sigma_{\rm NOISE}$ errors of the intensities in
Fig.\,\ref{fig8} are smaller than the height of the square symbols and
can therefore be neglected.

Similarly, we estimated the noise level of the spatially-averaged
intensities in Fig.\,\ref{fig9} from $1\sigma$ of the spatial
variations of the intensity over the selected area. In other words, we
assumed that the per-pixel noise level within the area is $1\sigma$ of
intensity variations over the area. However, the spatial averaging of
the intensities reduces their noise level by a factor of $1/\sqrt N$
or $1/61$, where $N$ is the number of averaged pixels. This again
points out, that the errors of intensities in Fig.\,\ref{fig9} are
much smaller than the height of the square symbols and are thus
insignificant.

In this paper, we define the reference wavelength $\lambda_{\rm R}$, or
zero point of the core velocity, as the spatio-temporal average of the
wavelengths $\lambda_{\rm C}$ of the core fit minima determined at
each pixel. We estimated uncertainty or stability of $\lambda_{\rm R}$
during the 71-min observing period in two ways. First, we constructed
71 proxy \ha\ profiles from instantaneous spatially-averaged
intensities, determined the wavelengths of their core fit minima, and
computed the median of their absolute differences from the total
average. Second, we took 71 spatial averages of the wavelengths of
core fit minima for each moment of the observation and computed again the
median of their absolute differences from the total average. Both
estimates gave consistent results, indicating that uncertainty of the
reference wavelength $\lambda_{\rm R}$ of the core velocity is about
$\pm 0.3$\kms.

When determining the reference wavelength $\lambda_{\rm R}$ we made no
distinction between the network and internetwork areas and this might be a
source of a systematic offset. We checked this feature by selecting only large
internetwork areas for determination of $\lambda_{\rm R}$. This one is
blueshifted about 0.2\kms\ with respect to the reference wavelength
from the whole field of view.

\section{Results}
\label{results}

This section presents a pictorial overview of results inferred from
the data undergone various processing. Gray shades in
Figs.\,\ref{fig10} -- \ref{fig12} highlight differences between the
uncorrected (gray) and corrected (black) data, while the different
styles distinguish the results of intensity averaging (solid lines)
and spatial averaging of spectral characteristics (symbols) over the
network element.

\begin{figure}
  \centering
  \includegraphics[width=\textwidth, clip=]{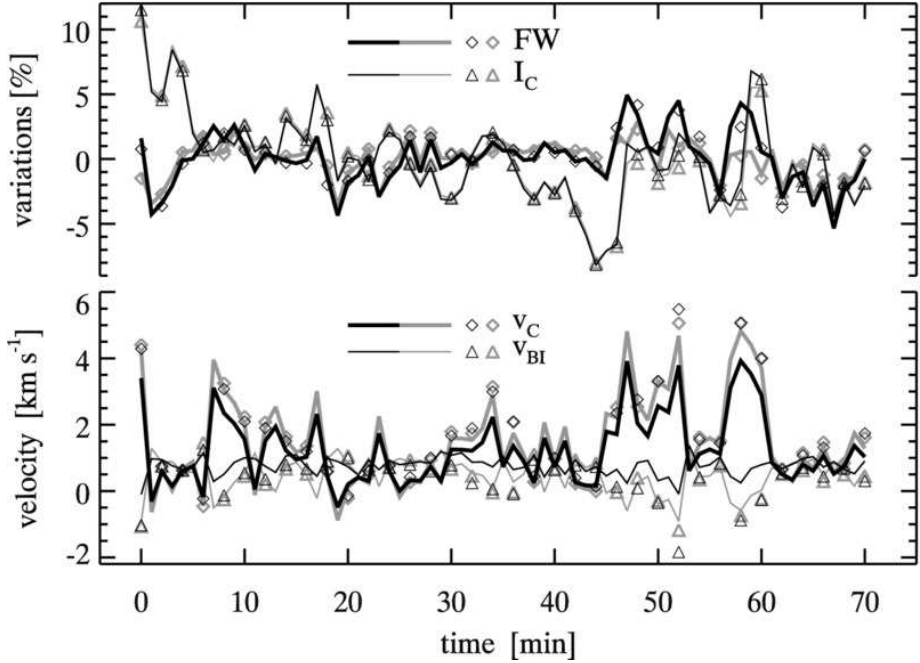}
  \caption{Temporal evolution of the fit width \fw, the core intensity
    \ic, the core velocity \vc, and the bisector velocity \vbi\ of the
    line core fit of the \ha\ spectral line in the network element
    (Figs.\,\ref{fig4}--\ref{fig7}) derived by the $4^{\rm th}$-order
    polynomial. \fw\ and \ic\ are shown as variations with respect to
    their temporal means (top panel). Positive \vc\ and
    \vbi\ signalize a redshift (middle and bottom panel). Gray and black
    lines indicate the uncorrected and corrected characteristics,
    respectively, derived from spatial means of the intensities. Gray
    and black symbols shown with the time step of 2\,min indicate the
    uncorrected and corrected characteristics, respectively, derived
    as spatial means of particular values at each pixel.}
  \label{fig10}
\end{figure}

\subsection{Temporal evolution of the spectral characteristics}

The top panel of Fig.\,\ref{fig10} shows temporal evolutions of the
relative variations of the fit width \fw\ and the core intensity
\ic\ of the line core fits of the \ha\ spectral line in the network
element (Figs.\,\ref{fig4}--\ref{fig7}) with respect to their temporal
means. The bottom panel shows temporal evolutions of the core velocity
\vc\ and the bisector velocity \vbi, whose positive values indicate
redshift and redward asymmetry of the line core. All these
characteristics were derived from the original \ha\ datacubes
downloaded from the DOT
database\footnote{\url{http://dotdb.strw.leidenuniv.nl/DOT/}}. An
application of the corrections (\pap) increases apparently the
amplitude of the \fw\ variations after the $40^{\rm th}$\,min, but
decreases both velocities within the whole 71-min period. The values
of \fw\ and \ic\ seem to be largely insensitive to the choice of
averaging for the size of the selected area, but this is not the case
of \vc\ and \vbi. Their corrected values resulting from spatial
averaging over the network element (black diamonds and triangles in
the bottom panel) differ systematically from the values inferred from
spatial means of the intensities (black thick and thin lines). This is
probably due to high non-linearity of the respective correcting
functions shown in \pap. The temporal evolutions of the
characteristics do not show any apparent systematic trend and fast
short-period variations dominate them. The variations of corrected
\fw\ are within the range $\pm 5$\,\% around the mean. The variations
of \ic\ indicate a rapid darkening of the selected network element
(Fig.\,\ref{fig4}) from 11\,\% above an average to average within the
first 10 minutes. Later on, \ic\ variations lie within the range
from $-9$\,\% to $+6$\,\%. The corrected \vc\ displays redshifts up to
4\kms, with an average of 1.5\kms\ and a noticeable correlation with
the corrected \fw, most apparent after the fortieth minute. In the
same period, a large expanding granule with the maximum diameter
larger than 2\,arcsec appeared in the low photosphere (see
Figs.\,\ref{fig6} and \ref{fig7}). Fig.\,\ref{fig10} suggests also
some similarity between the \vc\ and \ic\ variations. The Fourier
cross correlation indicates that \ic\ lags behind \vc\ about 2.1\,min
on average. The \ha\ core possesses mostly a redward asymmetry
corresponding to an inverse-C bisector with the average bisector
velocity of about 0.5\kms\ with excursions up to 1\kms.

\begin{figure}
  \centering
  \includegraphics[width=0.85\textwidth, clip=]{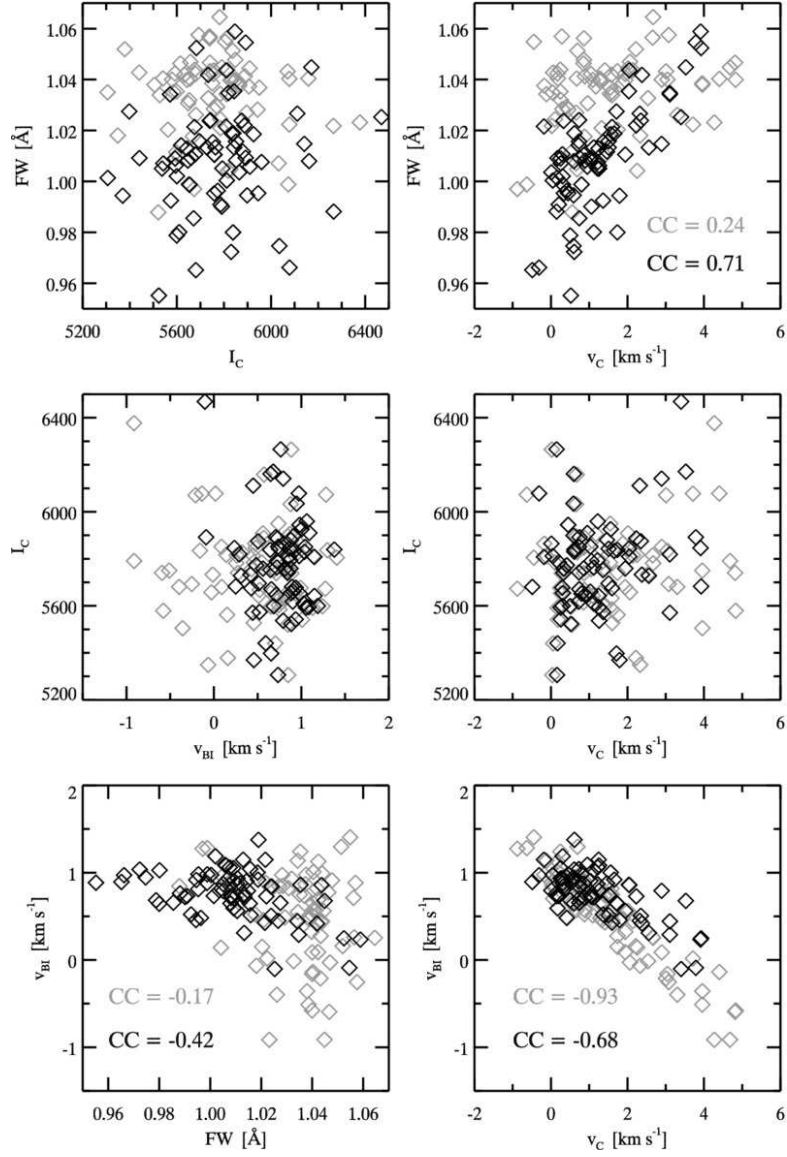}
  \caption{Scatter plots of the spectral characteristics shown in
    Fig.\,\ref{fig10} -- the case of spatial means of the
    intensities. Gray and black symbols indicate the uncorrected and 
    corrected characteristics, respectively. Unlike Fig.\,\ref{fig10},
    the plots show absolute values of \fw\ and \ic. Positive \vc\ and
    \vbi\ signalize a redshift. There are also shown the correlation
    coefficients CC for the pairs \fw\ -- \vc, \vbi\ -- \fw, and
    \vbi\ -- \vc\ for uncorrected (gray) and the corrected (black)
    values.}
  \label{fig11}
\end{figure}

\begin{figure}
  \centering
  \includegraphics[width=0.85\textwidth, clip=]{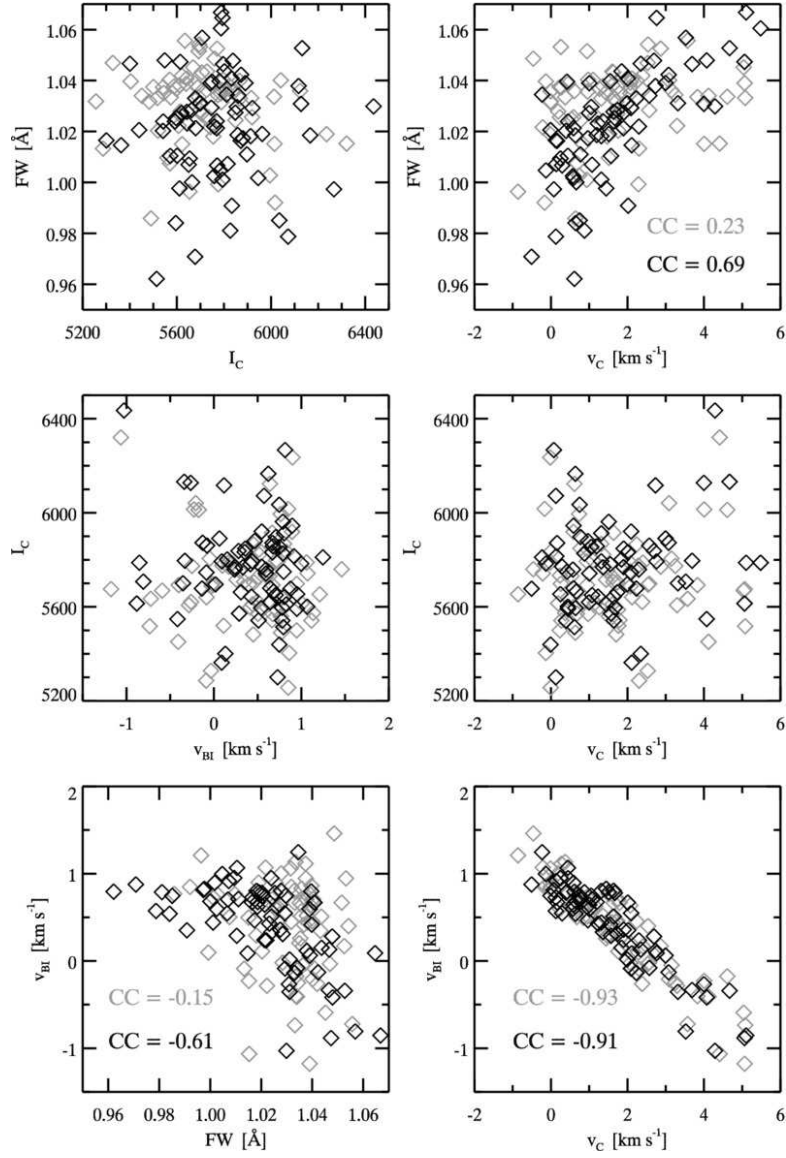}
  \caption{Scatter plots of the spectral characteristics shown in
    Fig.\,\ref{fig10} -- the case of spatial means of the spectral
    characteristics at each pixel.  Gray and black symbols indicate
    the uncorrected and corrected characteristics,
    respectively. Unlike Fig.\,\ref{fig10}, the plots show absolute
    values of \fw\ and \ic. Positive \vc\ and \vbi\ signalize a
    redshift. There are also shown the correlation coefficients CC for
    the pairs \fw\ -- \vc, \vbi\ -- \fw, and \vbi\ -- \vc\ for the
    uncorrected (gray) and corrected (black) values.}
  \label{fig12}
\end{figure}

\subsection{Scatter plots}

The spectral characteristics from Fig.\,\ref{fig10} are displayed in
the form of scatter plots in Figs.\,\ref{fig11} and \ref{fig12},
showing separately the results of spatial averaging of the intensities
and the spectral characteristics at each pixel, respectively. The top
left panels show \fw\ against \ic\ which do not correlate, neither
uncorrected nor corrected. The top right panels suggest a positive
correlation between \fw\ and \vc, which occurs only for the
corrected values. The correlation implies that more redshifted
profiles tend to be wider. The middle panels indicate that there is no
apparent correspondence between \ic\ and the velocities. The values of
\ic\ do not correlate with \vc, not even after shifting \ic\ forward
({\it i.e.}, to the left with respect to the zero of the time axis)
about 2.1\,min to compensate its apparent lag indicated in
Fig.\,\ref{fig10}. The bottom left panel of Fig.\,\ref{fig12} suggests
an anticorrelation between \vbi\ and \fw, which occurs only for the
corrected values resulting from their spatial averaging at each
pixel. The anticorrelation implies that while narrower \ha\ cores
display a redward asymmetry, the wider and more redshifted profiles (see
the \fw\ -- \vc\ correlation above) display a blueward
asymmetry. The bottom right panels also suggest a striking
anticorrelation between \vbi\ and \vc. Notably, the profiles with the
zero Dopplershift display a redward asymmetry with \vbi\ = 1\kms\ and
become symmetric for \vc\ of 2\kms, but tend to display an increasing
blueward asymmetry with further-increasing redshift.

\section{Discussion}

We derived spectral characteristics of the \ha\ spectral line observed
in the network element with the aim of seeking signatures of the
Alfv\'{e}n waves as reported in Jess {\it et al.} (2009). How do our
findings compare to the results of these authors? While Jess {\it et
  al.}  (2009) reported an average blueshift of 23\kms\ in a cluster
of bright points seen in \ha, we found an average redshift of about
1.5\kms. Obviously, this is a large discrepancy and we suspect that
our particular choice of the reference wavelength of the core velocity
might be partially responsible for that. We remind that we made no
distinction between the network and internetwork (see
Section\,\ref{datanal}). But as we showed in Section\,\ref{errors},
selecting only internetwork areas shifts the reference wavelength
blueward only insignificantly up to 0.2\kms\ and the value of the
average redshift is significantly above the uncertainty of 0.3\kms\ of
the reference wavelength. Further, Jess {\it et al.} (2009) reported
an absence of any significant intensity variations in the cluster of
bright points. However, there is some uncertainty whether they refer
to the spectral intensity in the line core, or the integrated
intensity. Our results suggest that the core width variations in the
selected network element are associated with both the core velocity
and core intensity variations. Finally, the authors detected {\it
  FWHM} oscillations of the \ha\ spectral line with the strongest
power in the 400-to-500\,s interval. We postpone detailed cross
correlation, Fourier, and wavelet analyzes of our results to a
forthcoming paper. We conclude that, most likely, a different
mechanism works in the selected network element than in the cluster of
the bright points studied by Jess {\it et al.}  (2009). The inverse-C
bisector and the average redshift of the \ha\ line core are more
symptomatic for propagation of chromospheric shocks, as suggested in
Uitenbroek (2006) and demonstrated in numerical simulations by
Heggland {\it et al.}  (2011). Then the lagging of the core intensity
\ic\ behind the core velocity \vc\ variations about 2.1\,min might be
characteristic either for the {\it p}-mode oscillations leaking into
the chromosphere, or for the shock wave propagation. Although there is
an apparent similarity of their variations best seen after the
fortieth minute (Fig.\,\ref{fig10}), shifting \ic\ forward about the
2-min lag does not improve the \ic\ -- \vc\ correlations shown in the
middle right panels of the scatter plots in Figs.\,\ref{fig11} and
\ref{fig12}, which may be ascribed to a variable phase shift which varies
in time and/or frequency.

Since the standard DOT observations do not involve any precise
wavelength calibration of the \ha\ filtergrams, we refrained from
interpreting straightforwardly redshifts and blueshifts as actual mass
downflows and upflows, respectively. Despite these facts, we will try
to reason that most of the positive values of \vc\ indicated by the
redshifts of the \ha\ core may represent the actual mass downflows. As
we noted in Section\,\ref{obs}, the zero wavelength of the filtergrams
is centered on the minimum of an average profile taken before the
observations over the field of view. Thus, the zero wavelength
approximately corresponds to the \ha\ solar disc-center wavelength of
$\lambda_{\rm sol} = 6562.81$\,\AA\ according to the spectral FTS
atlas by Neckel (1999), whose wavelength scale was already corrected
for the rotational and radial velocity of the Earth. Consequently, the
reference wavelength given in Section\,\ref{datanal} corresponds to
$\lambda_{\rm R} = \lambda_{\rm sol} - 0.018$\,\AA\ =
6562.79\,\AA. The local standard of rest at the solar surface for the
\ha\ spectral line is: $\lambda_0 = \lambda_{\rm lab}(1 + v_{\rm
  GR}/c) = 6562.80$\,\AA, where $\lambda_{\rm lab} = 6562.79$\,\AA\ is
the air laboratory wavelength of \ha\ adopted from the NIST database
(Kramida {\it et. al}, 2012), $v_{\rm GR} = 0.636$\kms\ is the solar
gravitational redshift, and $c$ is the speed of light. Then the core
velocities of \ha\ measured within this study are offset only about
$(\lambda_0 - \lambda_{\rm R})c/\lambda_0=0.5$\kms\ with respect to
the local standard of rest at the target area, with the following
consequences:
\begin{itemize} \itemsep=-0.5ex
\item the zero of \vc\ and all values of \vbi\ in Fig.\,\ref{fig10}
  should be shifted redward ({\it i.e.}, up) about 0.5\kms,
\item the zeros of the \vc\ axes in Figs.\,\ref{fig11} and \ref{fig12}
  should be shifted redward ({\it i.e.}, to the right) about 0.5\kms,
\item most of the positive values of \vc\ indicated by the redshifts
  of the \ha\ core may represent the actual mass downflows.
\end{itemize} 

\section{Conclusions}

A selected network element exhibits distinct spectral characteristics
seen in their temporal evolutions and scatter plots correlating pairs
of values from the same instant. In general, this involves:
\begin{itemize} \itemsep=-0.5ex
\item the fit width variations about $\pm 5$\,\%, {\it i.e.}, a few
  tens of milli{\aa}ngstr\"oms with respect to the average of
  1.01\,\AA,
\item the core intensity variations about $\pm 10$\,\% with respect to
  the average,
\item the core velocity variations from 0 to 4\kms\ about the average
  of 1.5\kms, implying a redshift,
\item the core asymmetry variations of the \ha\ line core up to
  1\kms\ about the average of 0.5\kms, suggesting an
  inverse-C bisector.
\end{itemize}
The \ha\ core width tends to correlate with the Dopplershift and
anticorrelate with the asymmetry, suggesting that more redshifted
\ha\ profiles are wider and the broadening of the \ha\ core is
accompanied with a change of the core asymmetry from redward to
blueward. We found also a striking anticorrelation between the core
asymmetry and the Dopplershift, suggesting also a change of the core
asymmetry from redward to blueward, with an increasing redshift of the
\ha\ core. A question is which of these patterns are unique just for
the network and which are also common in the internetwork. This can be
resolved in a stand-alone study, comparing the spectral
characteristics of the \ha\ line in the network and
internetwork. Finally, an absence of blueshift and detected intensity
and velocity variations probably exclude a presence of Alfv\'{e}n
waves in the selected network element according to the criteria given
in Jess {\it et al.} (2009).

\acknowledgements
 This work was supported by the Slovak Research and Development Agency
 under the contract No. APVV-0816-11. This work was supported by the
 Science Grant Agency - project VEGA 2/0108/12. This article was
 supported by the realization of the Project ITMS No. 26220120029,
 based on the supporting operational Research and development program
 financed from the European Regional Development Fund. The Technology
 Foundation STW in the Netherlands financially supported the
 development and construction of the DOT and follow-up technical
 developments. The DOT has been built by instrumentation groups of
 Utrecht University and Delft University (DEMO) and several firms with
 specialized tasks. The DOT is located at Observatorio del Roque de
 los Muchachos (ORM) of Instituto de Astrof\'{\i}sica de Canarias
 (IAC). DOT observations on 19 October 2005 have been funded by the
 OPTICON Trans-national Access Programme and by the ESMN-European
 Solar Magnetic Network - both programs of the EU FP6.

\end{document}